\documentstyle[twoside,12pt]{article}

\catcode`\@=11
\long\def\@makefntext#1{ 
\protect\noindent \hbox to 3.2pt {\hskip-.9pt
$^{{\eightrm\@thefnmark}}$\hfil}#1\hfill} 

 \def\@makefnmark{\hbox to 0pt{$^{\@thefnmark}$\hss}}  

\def\ps@myheadings{\let\@mkboth\@gobbletwo
\def\@oddhead{\hbox{} 
\rightmark\hfil\eightrm\thepage}
\def\@oddfoot{}\def\@evenhead{\eightrm\thepage\hfil 
\leftmark\hbox{}}\def\@evenfoot{}
\def\sectionmark##1{}\def\subsectionmark##1{}}

\oddsidemargin=\evensidemargin
\newcounter{sectionc}\newcounter{subsectionc}\newcounter{subsubsectionc}
\renewcommand{\section}[1] {\vspace{12pt}\addtocounter{sectionc}{1}
\setcounter{subsectionc}{0}\setcounter{subsubsectionc}{0}\noindent
	{\bf\thesectionc. #1}\par\vspace{5pt}}
\renewcommand{\subsection}[1] {\vspace{12pt}\addtocounter{subsectionc}{1}
	\setcounter{subsubsectionc}{0}\noindent
	{\bf\thesectionc.\thesubsectionc. {\kern1pt \bf\it #1}}\par\vspace{5pt}}
\renewcommand{\subsubsection}[1] {\vspace{12pt}\addtocounter{subsubsectionc}{1}
	\noindent{\thesectionc.\thesubsectionc.\thesubsubsectionc.
	{\kern1pt \it #1}}\par\vspace{5pt}}
\newcommand{\nonumsection}[1] {\vspace{12pt}\noindent{\bf #1}
	\par\vspace{5pt}}

\newcommand{\textlineskip}{\baselineskip=14pt}
\newcommand{\smalllineskip}{\baselineskip=12pt}

\def\eightcirc{
\begin{picture}(0,0)
\put(4.4,1.8){\circle{6.5}}
\end{picture}}
\def\eightcopyright{\eightcirc\kern2.7pt\hbox{\eightrm c}}

\newcounter{itemlistc}
\newcounter{romanlistc}
\newcounter{alphlistc}
\newcounter{arabiclistc}

\newcommand{\fcaption}[1]{
        \addtocounter{figure}{1}
         {{\tenrm Fig.~\thefigure . #1} }\hfil\break }

\newcommand{\tcaption}[1]{			
        \addtocounter{table}{1}
         {{\tenrm\offinterlineskip Table~\thetable . #1} }\hfil\break }

\def\thebibliography#1{\nonumsection{\large \bf References}\list
  {[\arabic{enumi}]}{\settowidth\labelwidth{[#1]}\leftmargin\labelwidth
    \advance\leftmargin\labelsep
    \usecounter{enumi}}
    \def\newblock{\hskip .11em plus .33em minus .07em}
    \sloppy\clubpenalty4000\widowpenalty4000}

\def\pmb#1{\setbox0=\hbox{#1}
	\kern-.025em\copy0\kern-\wd0
	\kern.05em\copy0\kern-\wd0
	\kern-.025em\raise.0433em\box0}

\def\fnt#1#2{\footnotetext{\kern-.3em
	{$^{\mbox{\scriptsize #1}}$}{#2}}}

\def\fpage#1{\begingroup
\voffset=.3in
\thispagestyle{empty}\begin{table}[b]\centerline{\footnotesize #1}
	\end{table}\endgroup}

 1
 1
 1

\font\eightrm=cmr8

\def\qed{\hbox{${\vcenter{\vbox{                          
   \hrule height 0.4pt\hbox{\vrule width 0.4pt height 6pt
   \kern5pt\vrule width 0.4pt}\hrule height 0.4pt}}}$}}

\newcommand{\be}{\begin{eqnarray}}
\newcommand{\ee}{\end{eqnarray}}
\newcommand{\dslash}{\partial \hskip -0.5em /}
\newcommand{\Dslash}{D \hskip -0.7em /}

\newcommand{\tr}{{\rm tr}}
\newcommand{\Tr}{{\rm Tr}}

\newcommand{\A}{{\cal A}}

\input psfig

\begin{document}
\normalsize\textlineskip
{\thispagestyle{empty}
\setcounter{page}{1}

\fpage{1}
\rightline{UNITU-THEP-7/1993}
\rightline{June 1993}
\vspace{1cm}
\centerline{\Large \bf The Nambu-Jona-Lasinio Chiral Soliton}
\vspace{0.5cm}
\centerline{\Large \bf with Constrained Baryon Number $^\dagger $}
\vspace{0.9cm}
\centerline{J. Schlienz, H. Weigel, H. Reinhardt and R. Alkofer}
\vspace{0.9cm}
\centerline{Institute for Theoretical Physics}
\vspace{0.3cm}
\centerline{T\"ubingen University}
\vspace{0.3cm}
\centerline{Auf der Morgenstelle 14}
\vspace{0.3cm}
\centerline{D-72076 T\"ubingen, FR Germany}

\vspace{3.0cm}
\normalsize\textlineskip
\noindent
\centerline{\bf Abstract}
\vspace{0.5cm}

A regularization for the baryon  number consistent with the energy in
the Nambu-Jona-Lasinio model is introduced. The soliton solution is
constructed with the regularized baryon number constrained to
unity. It is furthermore demonstrated that this constraint prevents
the soliton from collapsing when scalar fields are allowed to be space
dependent. In this scheme the scalar fields actually vanish at the origin
reflecting a partial restoration of chiral symmetry.
Also the influence of this constraint on some static properties
of baryons is discussed.

\noindent

\vfill

\noindent
$^\dagger $
{\footnotesize{Supported by the Deutsche Forschungsgemeinschaft (DFG) under
contract Re 856/2-1.}}
\eject

\normalsize\textlineskip
\nonumsection{\it Introduction}
In recent years the investigation of the solitons of the Nambu-Jona-Lasinio
(NJL) model \cite{na61} have experienced steady progress
\cite{re88,me89,al90,al92,al93}. While in the Skyrme model
\cite{sk61,an83} the baryon number is assumed to be given by the
topological charge \cite{wi83} the chiral soliton of the NJL model is not
strictly a topological soliton. On the contrary, the baryon number is given
more directly in terms of the quark fields and, in general, contains
contributions from the valence quarks as well as from the Dirac sea
\cite{re89}. In leading order of the derivative expansion the latter
is given by the topological charge \cite{eb86}. Whether the
baryon number is dominantly carried by the sea or the valence quarks
depends on the specific features of the model \cite{al92}.

Since the NJL action is not renormalizable regularization is required
and more importantly the resulting energy functional, baryon number
density, charge radii, etc. will depend on the regularization scheme employed
\cite{me90}.
Regularization is commonly applied to the Wick-rotated Euclidean
NJL action. A special feature of this Euclidean action is the fact that only
its real part is ultraviolet divergent while the imaginary part stays
finite. Accordingly in most treatments only the real part undergoes
regularization. As long as time components of vector and axialvector
fields are ignored only the real part contributes to the static energy
functional,
falsely pretending that the imaginary part is of no importance for
soliton solutions. The imaginary part is indeed relevant because it
completely determines the baryon number current. Since we intend to explore
the unit baryon number solutions the problem of
regularizing the imaginary part plays a central role.
Actually, a regularized imaginary part will not
{\it a priori} yield an integer
baryon number.

The aim of this letter is to investigate the soliton solutions of the
NJL model with a regularized imaginary part, {\it i.e.} a regularized
baryon number. Unit baryon number will be enforced by adding an appropriate
constraint to the energy functional. We will demonstrate that this
constraint cannot be satisfied by considering only the chiral angle field.
However, a unit regularized baryon number is attainable when additionally
scalar degrees of freedom are allowed to vary in space. This feature is
also related to the recently observed collapse of the NJL soliton
\cite{wa92}. It is unstable against building up a
narrow and infinitely high peak of the scalar field at the origin. The
valence quark is thereby joining the Dirac sea such that infinitely
many avoided crossings occur.  This effectively corresponds to a
situation where a level stemming from the positive part of the
spectrum acquires an infinitely large negative energy eigenvalue.
Without regularizing the baryon number such a configuration would
correspond to $B=1$. Obviously it should be clear that cutoff models
like the NJL model cannot be trusted when infinitely large energies
play a significant role. On the other hand, the regularized baryon
number tends  to zero during the collapse, {\it i.e.} baryon number is
leaking out of the soliton. In this letter we will demonstrate that
this leaking is prevented and the collapse is avoided
by fixing the
${\underline{\rm regularized}}$ baryon number to unity.
Then stable solutions exist with scalar
and pseudoscalar mesons included\footnote{The soliton with scalar and
pseudoscalar mesons is known to be also stabilized by adding a four
meson interaction \cite{wei93}.}. Furthermore we will discuss the
influence of fixing the regularized baryon number on several baryon
properties.

\medskip

\nonumsection{\it The model}
The starting point of our calculations is the two-flavor NJL action
$\A_{NJL}$ which, after bosonization \cite{eb86}, may be expressed as
the sum $\A_{NJL}=\A_F + A_m$ of a fermion determinant and a purely mesonic
part

\be
\A _F &=& \Tr \log (i\Dslash )
 = \Tr \log \big(i\dslash
- (P_R{\cal M}+P_L{\cal M^{\dag}}) \big),
\nonumber \\*
\A _m &=& \int d^4x \left( -\frac 1 {4g}
\tr ( {\cal M^{\dag}} {\cal M} - m_0({\cal M}+{\cal M^{\dag}}
) +m_0^2)   \right) .
\label{wirkung}
\ee
Here, $P_{R,L}$ are the projection operators on the right- and left-handed
quark fields, respectively, and
$m_0$= diag($m^u,m^d$) denotes the current quark mass matrix.
We will restrict ourselves to the two flavor case
and assume isospin symmetry: $m^u=m^d=m$.
The coupling constant $g$
will be determined from meson properties.
The complex field ${\cal M}=S+iP$ describes the scalar and pseudoscalar
meson fields which can be
parametrized by ${\cal M}=\Phi U$, $\Phi$ and $U$
being hermitian and unitary matrices, respectively. We will refer to
$\Phi$ as the chiral radius while the chiral angle $\Theta$ is
introduced via $U=\exp(i\Theta)$.

The quark determinant $\A_F$ diverges and must therefore be
regularized.
For the regularization procedure it is necessary to continue to Euclidean
space. This yields a complex Euclidean action
and we consider its real ($\A_R$)
and imaginary ($\A_I$) parts separately:
\be
\A_R &=& \frac 1 2 \Tr \log ( \Dslash_E ^{\dag}\Dslash_E )
\nonumber \\*
\A_I &=& \frac 1 2 \Tr \log ( (\Dslash_E ^{\dag})^{-1} \Dslash_E ) .
\ee
To keep the $O(4)$-invariance, we use
Schwinger's proper time regularization
\cite{sc51}.
For the real part this prescription consists of replacing
the logarithm by a parameter integral
\be
\A_R \rightarrow - \frac 1 2 \int_{1/\Lambda ^2}^\infty \frac {ds}s
\Tr \exp \left( -s \Dslash_E ^{\dag}\Dslash_E \right) ,
\ee
which for $\Lambda \to \infty $ reproduces the logarithm
up to an irrelevant constant.

Corresponding to the action (\ref{wirkung}) the energy of the static soliton
$E^{sol}=E^f+E^m$ splits into a fermionic and a mesonic part, where $E^f=
E^{val} +E^{vac}$ contains valence and vacuum parts.
To calculate $E^f$ we express the Euclidean Dirac operator
in its Hamiltonian form:
\be
i\Dslash_E=\beta(-\partial_{\tau}-h) ,
\label{vier}
\ee
wherein $\tau$ denotes the Euclidean time coordinate.
Substituting
the hedgehog ansatz
$U({\bf r})=\exp\left\{ i{\mbox{\boldmath $\tau$}}
\cdot\hat{{\bf r}}\Theta(r)\right\} $
for the chiral field and a radial function $\Phi=\Phi(r)$ for the
chiral radius
the static Hamiltonian reads:
\be
h={\mbox{\boldmath $\alpha$}}\cdot{\bf p}+\beta\Phi\left( \cos\Theta
+i\hat{{\bf r}}\cdot{\mbox{\boldmath $\tau$}}\gamma_5\sin\Theta\right) .
\label{fuenf}
\ee
Denoting the associated eigenvalues by $\epsilon_{\mu}$ the
fermionic contribution to the static energy is given by \cite{re89}:
\be
E^{vac}\left[\Theta,\Phi \right]
&=&\frac{N_c}{2}\frac{1}{\sqrt{4\pi}}\int_{1/\Lambda^2}^{\infty}
ds\,s^{-3/2}\sum_{\mu}\left( \exp\left(-s\epsilon_{\mu}^2\right)
-\exp\left(-s(\epsilon_{\mu}^0)^2\right) \right) \nonumber \\
E^{val}\left[\Theta,\Phi \right]
&=&N_c\sum_{\mu}\eta_{\mu}\left|\epsilon_{\mu}\right|
\label{hpro}
\ee
where $\eta_{\mu}$ denote the (anti-) quark occupation numbers and
$\epsilon_{\mu}^0$ are the eigenvalues of $h$ for
the mesonic vacuum ${\cal M}=
\langle {\cal M} \rangle =M{\underline{\rm 1}}$
with $M$ being the constituent quark mass. Note that the vacuum energy
has been subtracted in (\ref{hpro}).
The only free parameter is $M$ since
$\Lambda$ is fixed by fitting the pion decay constant
$f_{\pi}=$ 93MeV.
The gap equation relates the current quark mass $m$ to
the constituent quark mass $M$.
The coupling constant $g$ is eliminated by fitting the pion
mass $m_{\pi}=135$MeV: $g=mM/m_{\pi}^2 f_{\pi}^2$ \cite{eb86}.
Thus the mesonic part of the soliton energy may be expressed as
\be
E^m\left[\Theta,\Phi \right]=\frac{2\pi m_{\pi}^2 f_{\pi}^2}{mM}
\int dr\,r^2\,\left\{ \Phi^2(r)-M^2
-2m\left( \Phi(r)\cos\Theta(r)-M\right) \right\} .
\ee
Again the contribution of the vacuum configuration has been subtracted.

Until now, we did not take into account regularization of ${\cal A}_{I}$.
Since this part is finite, the question arises whether it has to be
regularized or not. Here we argue that regularization is necessary in
order to have a consistent appearance of the one particle eigenenergies
$\epsilon_{\mu}$ \cite{we92,zu93,al93,we93}.
$\A_I$ does not contribute to the soliton mass, however, its regularization
has drastic consequences when we regard the sea contribution to
baryon number
\be
B^{vac}=-\frac{1}{N_c} \lim_{T\to\infty} \frac{1}{T}\mbox{Tr}
\left\{ h\,(-\partial_{\tau}^2+h^2)^{-1}\right\} ,
\label{nine}
\ee
where $T$ is the Euclidean time interval under consideration.
We introduce regularization of the baryon number by replacing
\cite{al93,we93}
\be
 (-\partial_{\tau}^2+h^2)^{-1}
\rightarrow \int_{1/\Lambda^2}^{\infty}ds\,e^{-s(
-\partial_{\tau}^2+h^2)},
\ee
which again is an identity for $\Lambda\to\infty$.
The regularized baryon number $B_{\Lambda}$ then reads:
\be
B_{\Lambda}=\sum_{\mu}\mbox{sign}(\epsilon_{\mu})
{\cal N_{\mu}}
+\eta^{val} ,
\ee
wherein  ${\cal N_{\mu}}=-\frac{1}{2}\mbox{erfc}
\left( \left| \frac{\epsilon_{\mu}}{\Lambda}\right| \right) $
denote the vacuum occupation numbers of the single quark orbits $\mu$ in
the proper-time regularization \cite{re88}.
In the limit $\Lambda\to\infty,\quad B_{\Lambda}$ is obviously
integer, however, this is no longer the case for finite $\Lambda$.
The main purpose of this paper is to investigate self-consistent solitons
constrained to the baryon number $B_{\Lambda}=1$.
Accordingly, the energy functional is  replaced by
\be
E=E^{val}+E^{vac}+E^m+
M\left[\lambda(B_{\Lambda}-1)^2-\frac{1}{2}a\lambda^2 \right] .
\label{zehn}
\ee
The constituent quark mass $M$ is included such that the Lagrange multiplier
$\lambda$ and the auxiliary parameter $a$ are dimensionless.
Varying the total energy functional E with respect to $\lambda$ yields
\be
\lambda=\frac{(B_{\Lambda}-1)^2}{a} .
\label{dz}
\ee
The limit $a\rightarrow 0$ has to be assumed
in order to achieve $B_{\Lambda}=1$.
Variation of $E$ with respect to the meson fields provides
the equations of motion
\be
P(r)\cos\Theta(r)&=&\left[ S(r)-\frac{4\pi m_{\pi}^2 f_{\pi}^2}
{N_C M} \right] \sin\Theta(r) \nonumber\\
\Phi(r)&=&m\cos\Theta(r)-\frac{m N_C M}{4\pi m_{\pi}^2 f_{\pi}^2}
\left[ S(r)\cos\Theta(r)+P(r)\sin\Theta(r)\right]
\label{elf}
\ee
where we have denoted
\be
S(r)&=&\sum_{\mu} \left[ (\cal N_{\mu}+\eta_{\mu})\mbox{
sign}(\epsilon_{\mu})
+C_{\mu} \right] \int d\Omega\,\bar \Psi_{\mu}\Psi_{\mu} ,\nonumber\\
P(r)&=&\sum_{\mu} \left[ (\cal N_{\mu}+\eta_{\mu})\mbox{
sign}(\epsilon_{\mu})+C_{\mu} \right] \int d\Omega\,\bar \Psi_{\mu}
\left[ i{\mbox{\boldmath $\tau$}}
\cdot{\bf r}\Theta(r)\gamma_5 \right] \Psi_{\mu} .
\ee
The derivative of the constraint with respect to the one particle
eigenenergy $\epsilon_{\mu}$ is proportional to
\be
{\cal C_{\mu}}&=&\frac{2M\lambda(B_{\Lambda}-1)}{N_C\sqrt{\pi}\Lambda}
\exp\left\{-\frac{\epsilon_{\mu}^2}{\Lambda^2}\right\} .
\ee

\nonumsection{\it Numerical Results}
The self-consistent soliton is constructed by iterating
the constraint (\ref{dz}) and the equations
of motion (\ref{elf})
together with the corresponding eigenvalue problem
associated with the Hamiltonian (\ref{fuenf}). The latter is solved by
discretizing the momentum eigenvalues of the free Hamiltonian.
This is achieved by putting the system into a large spherical box
of radius D. Our calculations are performed with a constituent
quark mass $M=400$ MeV. We consider this to be sufficient in order
to demonstrate that the soliton is stabilized by constraining the
regularized baryon number to unity.

Restricting the meson profiles to the chiral circle and omitting
the constraint (\ref{zehn}) we find $B_{\Lambda}=0.966$ for
the regularized baryon number. This number cannot be augmented
considerably by including (\ref{zehn}) and staying on the chiral circle.
{\it I.e.} we do not find stable solutions in the limit $a\to 0$
when the chiral angle $\Theta$ is taken to be the only space dependent
meson field. Allowing, in addition, the chiral radius to be space
dependent as well, we face the well-known problem that the soliton of
scalar and pseudoscalar fields collapses. As discussed in the introduction
this collapse is associated with the leakage of baryon number
$B_{\Lambda}\to 0$ and as we will see,
including the constraint (\ref{zehn})
stabilizes the soliton. Indeed we find stable solutions
with scalar and pseudoscalar fields when the regularized baryon number
is constrained to unity. As can be observed from table 1, $B_{\Lambda}=1$
cannot exactly be fulfilled for finite $D$ and we conjecture that
$B_{\Lambda}=1$ is only attainable in the continuum limit, $D\to\infty$.

In table 2 we display the energies of the self-consistent soliton
solutions for various strengths of the constraint. These strengths
are given in terms of the auxiliary parameter $a$ and can be transferred to
different regularized baryon numbers  $B_{\Lambda}$.
One sees
that the valence quark eigenenergy increases as $B_{\Lambda}$
tends to unity.
The resulting meson profiles are still localized as
is obvious from figs. 1 and 2 where we display the meson profiles
$\Theta (r)$ and $\Phi(r)$ corresponding to the regularized baryon numbers
$B_{\Lambda}$ of table 2. The chiral angle $\Theta(r)$ exhibits a
strong squeezing at the origin with growing baryon number, while for
$r>1$ fm all profile functions show the same spatial dependence. Thus,
the constraint effects the meson profiles only at small $r$. This can also
be observed from the profile function $\Phi(r)$ which  almost vanishes at
the origin, reflecting a local partial restoration of chiral symmetry.
\begin{figure}
\centerline{\hskip -1.5cm
\psfig{figure=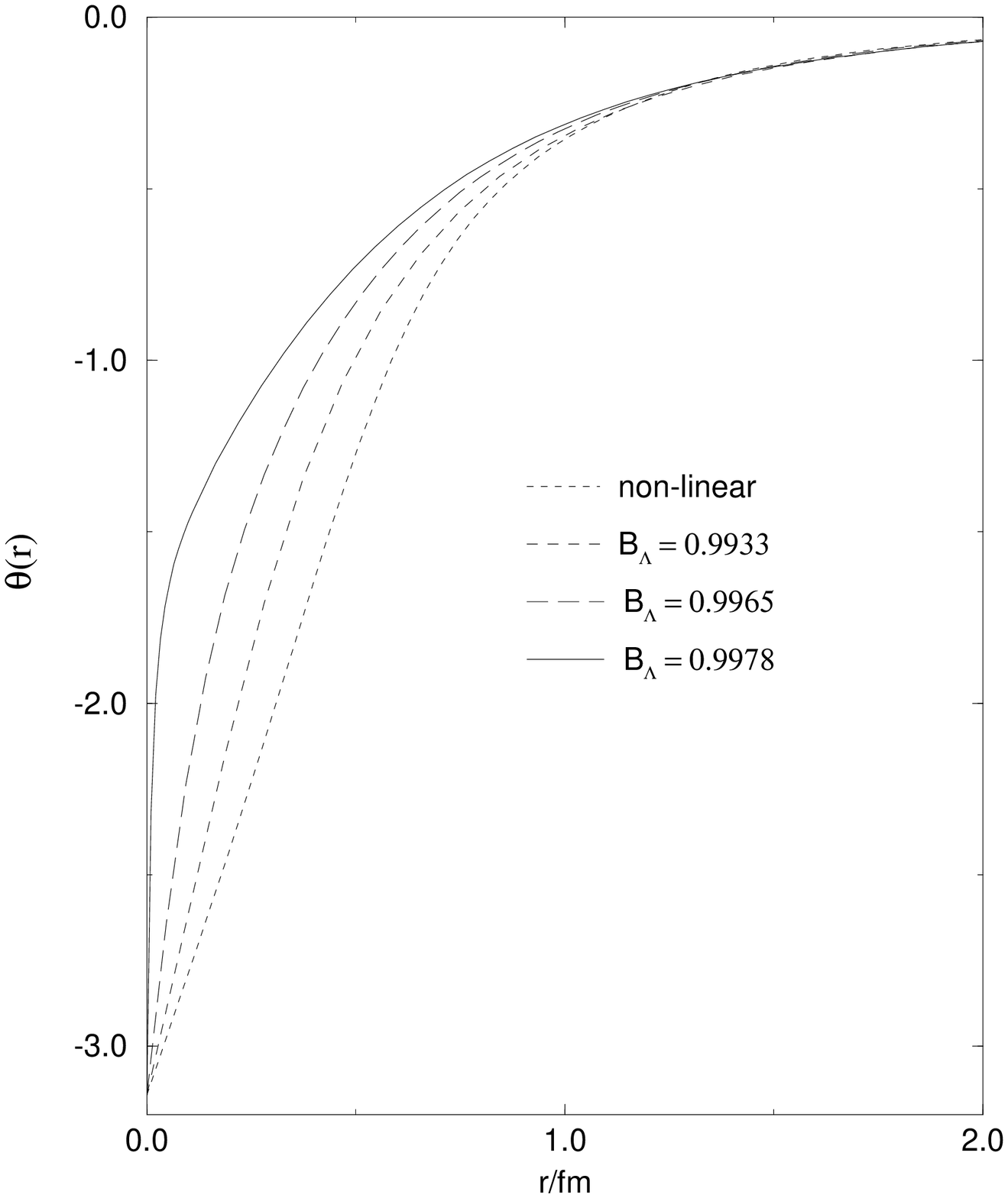,height=9.0cm,width=12.0cm}}
\vspace{5mm}
\fcaption{The chiral angle $\Theta(r)$ for the non-linear NJL
soliton (short dashed line) and different strengths of the constraint.
}
\end{figure}

\begin{figure}
\centerline{\hskip -1.5cm
\psfig{figure=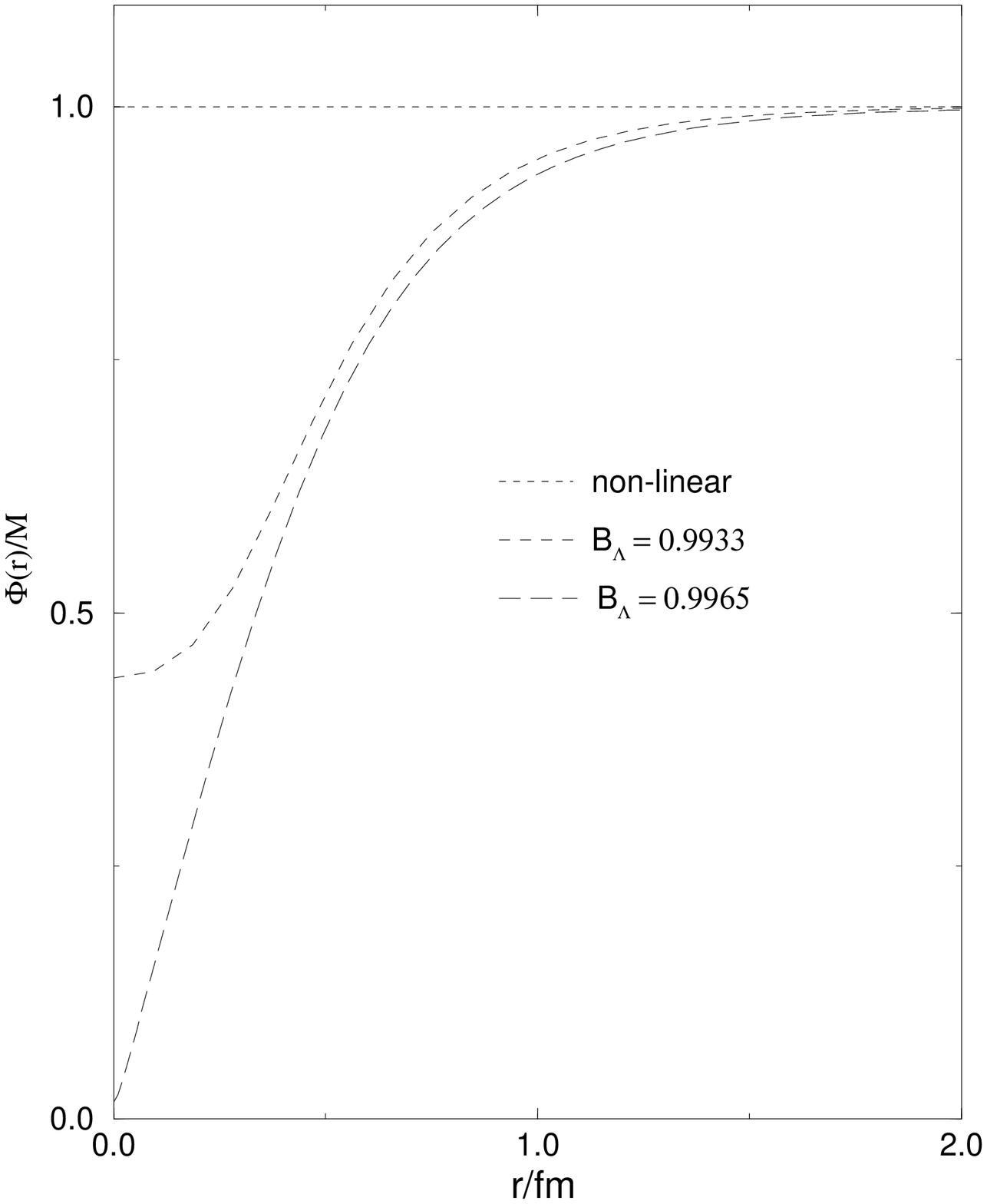,height=9.0cm,width=12.0cm}}
\vspace{5mm}
\fcaption{Same as fig. 1 for the chiral radius $\Phi(r)$.
}
\end{figure}

\begin{table}
\tcaption{The maximal baryon number obtained for different
box radii D.}
\newline
\centerline{\tenrm\smalllineskip
\begin{tabular}{||l|c||}
\hline
D/fm &$B_{\Lambda}$\\
\hline
4&0.9970\\
\hline
5&0.9976\\
\hline
6&0.9977\\
\hline
7&0.9979\\
\hline
8&0.9980\\
\hline
\end{tabular}}
\end{table}

\begin{table}
\tcaption{The soliton energy $E$ for constituent mass $M=$400 MeV
with its valence quark, sea
and mesonic contributions $E^{val}$, $E^{vac}$ and $E^m$ for different
baryon numbers $B_{\Lambda}$. The first line refers to the non--linear case.
The small deviation of $E$ from
$E^{val}+E^{vac}+E^m$ is due to the constraint in (11).
All energies are given in MeV.
}
\newline
\centerline{\tenrm\smalllineskip
\begin{tabular}{||l|c|c|c|c||}
\hline
$B_{\Lambda}$ & $E^{val}$ & $E^{vac}$ & $E^m$ & $E$   \\
\hline
0.9662 & 632 & 572 & 34 & 1238 \\
\hline
0.9932 & 914 & 753 & -387 & 1285  \\
\hline
0.9939 & 925 & 755 & -398 & 1286 \\
\hline
0.9946  & 941 & 758 & -426 & 1287  \\
\hline
0.9965  & 996 & 770 & -475 & 1293  \\
\hline
0.9973 & 1030 & 776 & -510 & 1294 \\
\hline
0.9978 & 1043 & 780 & -529 & 1296 \\
\hline
\end{tabular}}
\end{table}

In order to further investigate whether the solutions are localized
objects we have considered the baryon densities ({\it cf.} fig. 3).
Here, the sea quark part of the baryon density is of particular interest
because it almost vanishes, whereas the energy of the Dirac sea turns out
to be sizable. In order to understand this result, we split in fig. 4  the
sea quark density into two parts stemming from the intrinsic positive and
negative parity eigenfunctions of (\ref{fuenf}),
respectively. It can clearly be
seen that the contributions of these parts to the baryon density cancel.
For the static energy, however, they sum up coherently resulting in a
relatively large value.
\begin{figure}
\centerline{\hskip -1.5cm
\psfig{figure=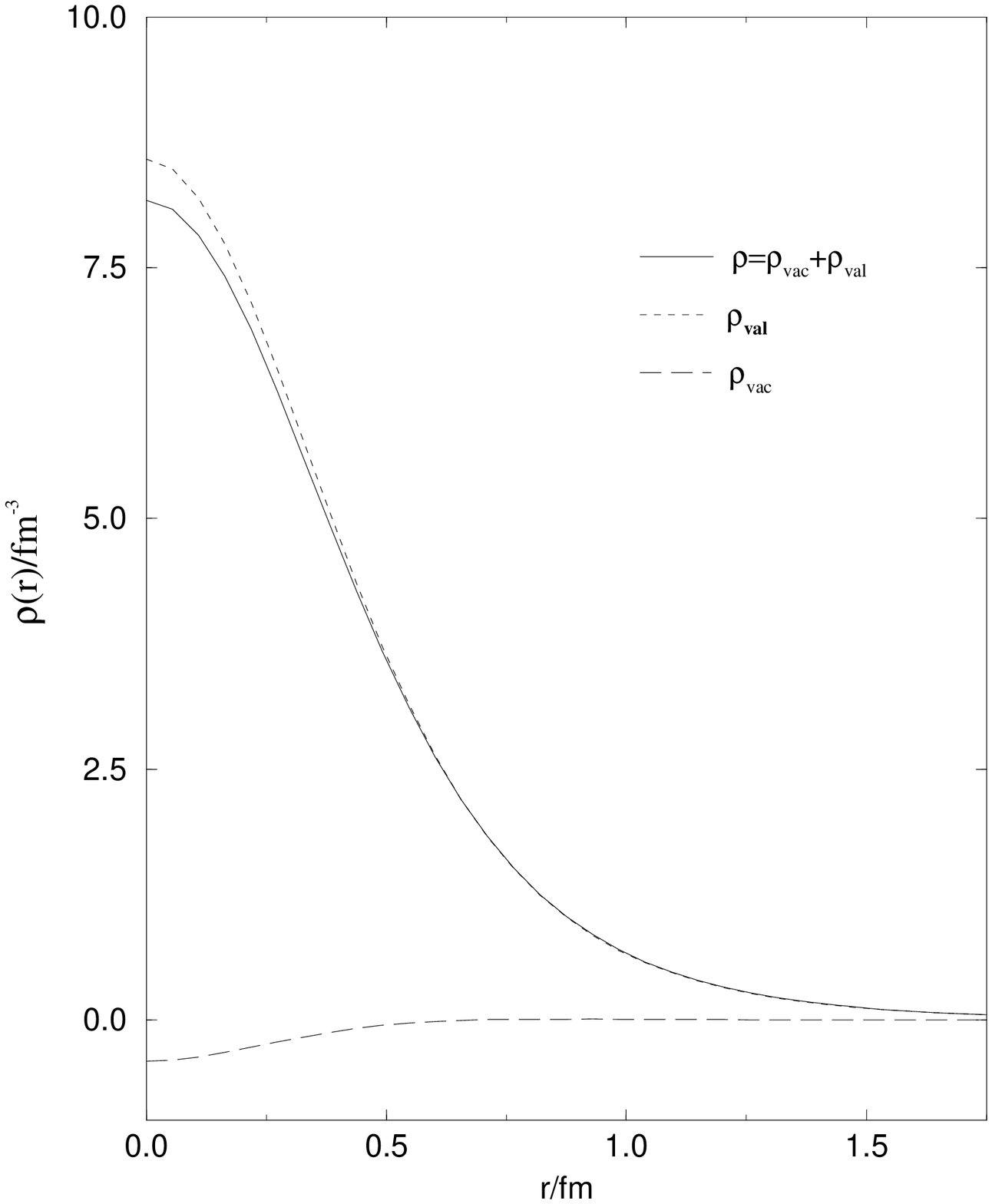,height=9.0cm,width=12.0cm}}
\vspace{5mm}
\fcaption{The baryon density $\rho(r)$ (full line) and its valence
(short dashed line) and sea quark contribution (long dashed line) for
the soliton with $B_{\Lambda}=0.9978$.
}
\end{figure}
\begin{figure}
\centerline{\hskip -1.5cm
\psfig{figure=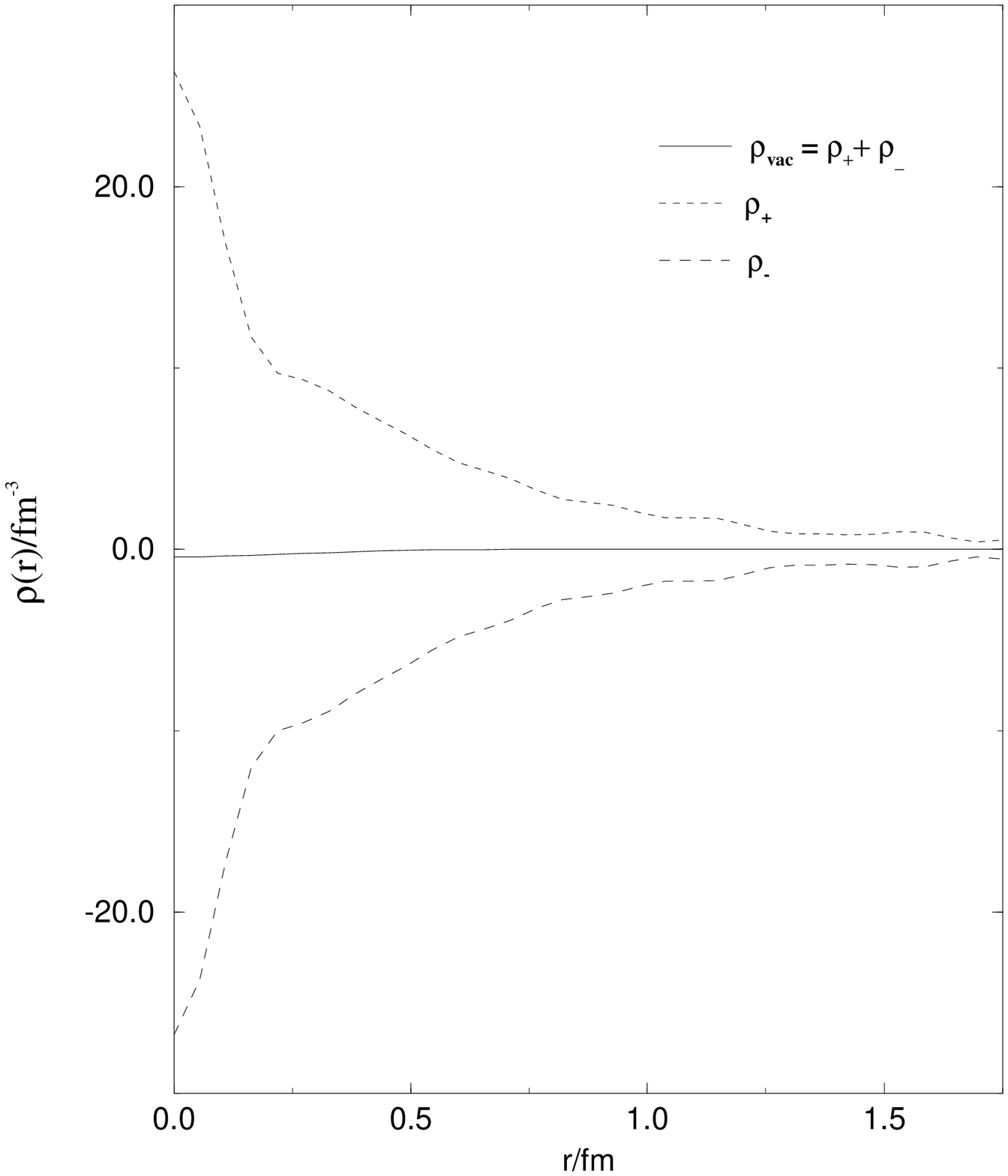,height=9.0cm,width=12.0cm}}
\vspace{5mm}
\fcaption{The sea quark contribution to the baryon density (full line)
and its contributions from states with intrinsic positive parity
(short dashed line) and negative intrinsic parity (long dashed line)
for the soliton with $B_{\Lambda}=0.9978$.
}
\end{figure}

\begin{table}
\tcaption{Baryon radius $\langle r_{I=0}^2 \rangle ^{1/2}$,
axial charge $g_A$ and the moment of inertia $\alpha ^2$
for different baryon numbers. The first line denotes the data
from the unconstrained soliton on the chiral circle ($\Phi\equiv M$).
}
\newline
\centerline{\tenrm\smalllineskip
\begin{tabular}{||l|c|c|c||}
\hline
$B_{\Lambda} $ & $\langle r^2_{I=0} \rangle^{1/2}$/fm
 & $\alpha^2/(\mbox{GeV})^{-1}$ & $g_A$ \\
\hline
0.9662 & 0.77 & 5.52 & 0.72 \\
\hline
0.9933 & 0.93 & 9.11 & 0.83 \\
\hline
0.9946 & 0.96 & 9.87 & 0.84 \\
\hline
0.9965 & 1.03 & 11.94& 0.86 \\
\hline
0.9973 & 1.08 & 13.54 & 0.87  \\
\hline
0.9978 & 1.11 & 14.62 & 0.89 \\
\hline
\end{tabular}}
\end{table}
Let us next turn to the discussion of a few static properties resulting
from our soliton solution. In table 3 we display the isoscalar radius
$\langle r^2_{I=0} \rangle^{1/2}$ which is associated to the baryon
charge distribution, the axial charge of the nucleon, $g_A$,
as well as the moment of inertia $\alpha^2$ for collective rotations
in isospace. The latter measures the $\Delta$-nucleon mass difference
$M_{\Delta}-M_n=3/2\alpha^2$. We should mention that the
analytical form of these quantities in terms of sums involving matrix
elements of the eigenstates of (\ref{fuenf}) do not differ from the
original treatment of the pure pseudoscalar case. We thus may employ the
relevant formulas from refs. \cite{re89,wa91}. In agreement with the
results for the static energy we observe for these quantities a dominating
valence quark contribution as $B_{\Lambda}$ tends towards unity.
Considering {\it e.g.} the moment of inertia, $\alpha^2$, the Dirac sea
contributions reduces from 23\% to 4\% when we go from the pure pseudoscalar
to $B_{\Lambda}=0.9978$. This scheme may easily be understood by remarking
that the quantities under consideration are rather sensitive to the
large $r$-behavior of the fields: The valence quarks get enhanced of
large $r$ while the meson fields are squeezed at the origin.
It is remarkable that the inclusions of the constraint improves
our predictions for $\langle r^2_{I=0} \rangle^{1/2}$ and $g_A$
while $\alpha^2\approx 15 \mbox{ GeV}^{-1}$ is far off the empirical value
$\alpha^2=3/2(M_{\Delta}-M_n)=5 \mbox{ GeV}^{-1}$.

\nonumsection{\it Conclusions}
The main result of this letter is that the collapse of the NJL soliton
is prevented when constraining the regularized baryon number to unity.
This constraint can only be fulfilled when the scalar meson field
is allowed
to be space dependent. Then it
displays a partial restoration of chiral symmetry close to the
origin. Despite of the fact that the valence quark
energy eigenvalue approaches
the one without soliton the valence quark wave function is still quite
well localized, and the Dirac sea is still polarized. The vacuum
contribution to the baryon number vanishes, however, the vacuum
contribution to the soliton energy is sizeable. Whereas some static
properties of the baryons, the isoscalar mean square radius and the
axial coupling $g_A$, are improved when considering the $B_{\Lambda}=1$
soliton, the moment of inertia is too large and therefore the predicted
nucleon-$\Delta$ splitting becomes too small.

We have seen that constraining the regularized
baryon number leads to a valence quark dominated picture of the
soliton. It would therefore be interesting to include the constraint
also in the case when vector and axial vector fields are present.
As is well known, these models support the Skyrmion picture of the
baryon, {\it i.e.} the valence quark joins the Dirac sea
\cite{al92}. First results indeed show that the inclusion of the
constraint leads again to an increased valence quark eigenenergy
which may even be positive.

In this letter we used the proper time regularization which already
``damps'' the contribution of low lying levels to physical quantities,
especially to the baryon number. From this point of view it is easily
understandable that levels are pushed ``out of the gap''. The question
remains whether a regularization procedure which does only influence
high-lying levels will lead to a different qualitative behavior. This
idea can probably be tested using dimensional regularization which is
known to even enhance contributions from low-lying levels \cite{kr92}.

\newpage

\end{document}